# Low-dimensional geometry learning for turbulence prediction in optimized stellarators

Xishuo Wei[1*], Handi Huang[1], Haotian Chen[1], Hongxuan Zhu[2,3], Zhe Bai[4], Samuel Williams[4], Zhihong Lin[1*]

1. University of California, Irvine, CA 92697, United States of America

2. Zhejiang University, Hangzhou, Zhejiang, China

2. Princeton University, Princeton 08540, NJ, United States of America

4. Lawrence Berkeley National Laboratory, Berkeley CA, United States of America

*Email: xishuow@uci.edu, zhihongl@uci.edu

## Abstract

Optimized stellarators offer improved energetic particle and neoclassical confinement, but optimization based on inexpensive proxies or local simulations may miss important global nonlinear physics. In this work, we investigate whether dimensionality reduction can enable data-efficient surrogate modeling of turbulence-relevant global properties in optimized stellarators. Using a database of more than 13,000 quasi-helical (QH) equilibria, we train an autoencoder and show that the geometric variation of QH configurations is well represented by a low-dimensional manifold. Based on this low-dimensional representation, we build surrogate models for two key quantities: the long-wavelength Rosenbluth–Hinton (RH) zonal-flow residual and the steady-state ion-temperature-gradient (ITG) turbulent heat flux from global gyrokinetic simulations. Both quantities exhibit organized structures in the learned latent space, enabling efficient exploration of the QH subspace and identification of low-transport configurations while preserving QH characteristics. We further identify a direct relationship between RH residual level and magnetic-axis excursion, consistent with near-axis expansion theory. Large-scale global gyrokinetic simulations confirm substantial transport variation across QH geometries, highlighting the value of latent-space-guided screening.

## 1. Introduction

Optimized stellarators are attractive fusion reactor concepts because they substantially improves energetic particle and neoclassical confinement in the 3D magnetic configuration[1], [2]. In practice, optimized stellarator configurations are typically obtained through numerical optimization, where a predefined objective function is minimized by iteratively adjusting design parameters that determine the equilibrium magnetic flux surfaces. Using this general strategy, previous studies have successfully constructed Quasi-symmetric (QS) and quasi-isodynamic (QI) configurations[2], [3], [4], [5], [6]. Optimization also commonly imposes basic shaping constraints (for example, aspect ratio and elongation), rotational transform ($\iota$) profile, and many additional physical objectives incorporated through computationally inexpensive proxies. Examples include Mercier criterion for MHD interchange stability [7], [8] and effective ripple as a proxy for neo-classical transport in low collisional regime[9], [10]. The recent W7-X experiments [11], [12] have confirmed that the neoclassical transport can be reduced through magnetic-field optimization, and the core energy confinement is determined by turbulence instead. These results motivate the turbulence optimization aimed at future reactors. Over the years, many results have used different proxies to minimize the linear drive and nonlinear transport of micro-instabilities.[5], [13], [14], [15], [16], [17], [18], [19], [20], [21], [22], [23], [24], [25]. Notably, Kim *et al* [26] has directly optimized the heat flux from local gyrokinetic simulations, while Landreman *et al* [27] trained the surrogate model from the local simulations to predict the heat flux, which can also be used for optimization.

A limitation of using the proxies and local simulations for optimization is that these simplified metrics may not capture the underlying global nonlinear behaviors including zonal flows[28], [29] and turbulence spreading [30], [31] in the 3D magnetic equilibrium. Furthermore, the recent nonlinear MHD simulations indicate that the linearly unstable ballooning activity in standard W7-X configuration can saturate benignly, consistent with the "soft" beta limit rather than a hard disruptive limit [32]. Regarding microturbulence, global gyrokinetic simulations in stellarators have reported regimes in which TEM excited by helically trapped electrons can drive substantial transport in the W7-X[31], and the different interaction between zonal flows and turbulence can significantly impact the steady-state transport level for different stellarator configurations[29], [33]. These findings highlight the importance of first-principle global simulations during optimization. However, these simulations are typically orders of magnitude more expensive than proxy evaluations, making them difficult to embed directly in the iterative optimization loop. Machine learning surrogate models, on the other hand, provide a potential path forward because they can deliver rapid inference and can be integrated with automatic differentiation in modern frameworks, enabling gradient-based design and optimization workflows. The key obstacle is the high dimensionality of parameter space in the optimization process. Stellarator optimization problems often involve hundreds to thousands of free parameters, which define a very high-dimensional design space. It is not feasible to populate this space with enough first-principle global simulations to train a globally reliable surrogate model. And even if a surrogate model is trained on a limited dataset, a fundamental question is why the surrogate can be expected to generalize for new configurations.

A useful observation is that physically meaningful stellarators occupy only a constrained subset because poor energetic particle and neoclassical confinement is unacceptable regardless of other performance metrics. Concepts such as QS and QI impose strong geometric constraints, which can restrict the relevant configurations to a small fraction of the nominal parameter space and can dramatically reduce the intrinsic dimensionality of the realizable designs. Related low dimensional structure has been reported in earlier stellarator design studies, for example in global searches over coil current settings [34]. Moreover, the near-axis-expansion theory itself is an approach to express the configurations approximately with few parameters. Finally, the internal kink mode structure in tokamak is found to have the low-dimensional feature and can be approximately described by 3 parameters [35]. These examples are consistent with the broader idea that high-dimensional datasets can concentrate near low-dimensional manifolds, which motives latent variable representations such as autoencoders[36]. In fact, the low dimensional feature is very common for naturally generated data, such as images and languages[37], [38], [39]

In this paper, we focus on one QS configuration, quasi-helical symmetric (QH) stellarator, to show that QH geometric distribution has an intrinsic dimension that is far smaller than the total number of free parameters used in the optimization process. Using a dataset of more than 13,000 QH equilibria, we train an autoencoder to reconstruct the geometry from a low-dimensional latent representation and find that a three-dimensional latent space is sufficient to capture the essential variation of the QH geometries. Using these latent representations, we construct surrogate models for two turbulence-relevant quantities, namely the zonal-flow residual level in the long wavelength limit [40][41], and the volume-averaged steady-state heat flux driven by ion temperature gradient (ITG) turbulence. We further show that both the zonal-flow residual and the simulated transport exhibit organized structures when projected onto the learned three-dimensional latent space, enabling efficient exploration of the QH subspace and the identification and generation of candidate low-transport configurations.

The remainder of this paper is organized as follows. Section 2 describes the data generation, the dimensionality-reduction method, and model architecture. Section 3 reports results for geometry reconstruction, zonal-flow residual prediction, and turbulent transport prediction. Section 4 summarizes the results and outlines future directions.

## 2. Data and Methods

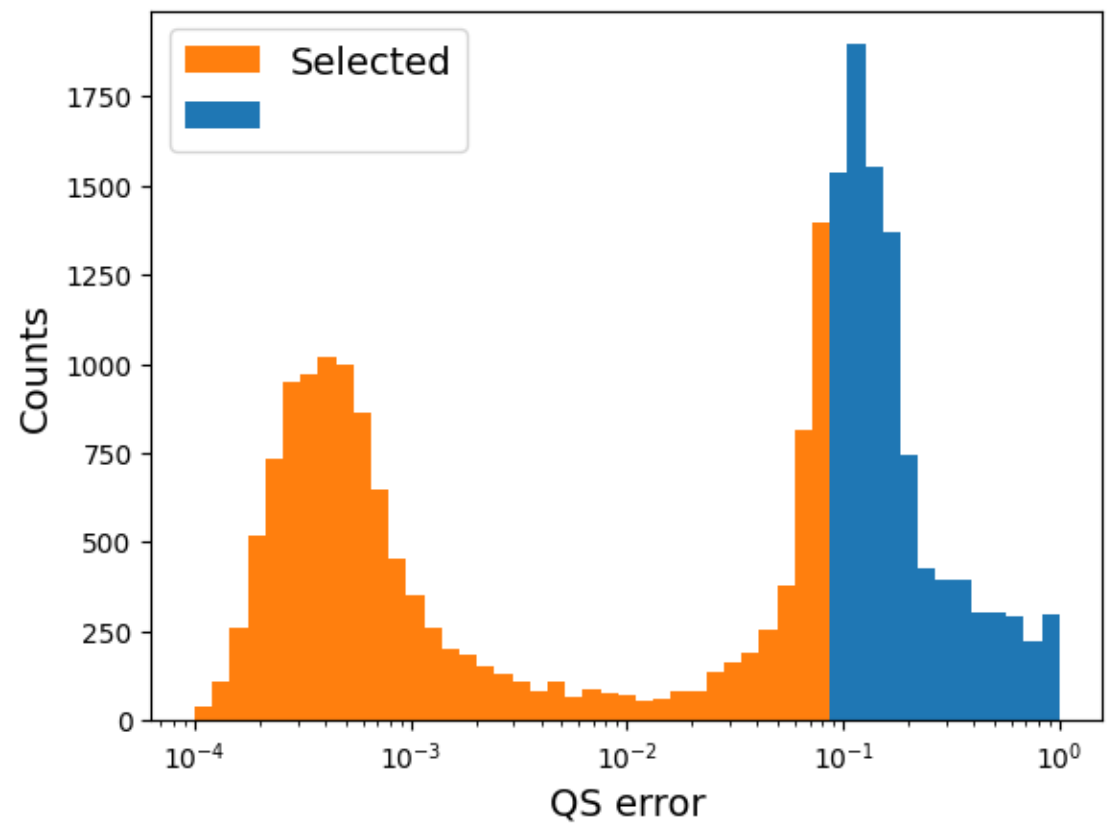

*Figure 1 Histogram of the QS error of the generated equilibria. The 13949 equilibria with QS error lower than 0.1 are selected to form the dataset.*

We used the optimization code DESC[42], [43] to generate a dataset of QH stellarator equilibria with toroidal period NFP=4. The initial parameters were selected to ensure sufficient diversity in geometry distribution. The initial flux surface cross section was set to be concentric circular, with the inverse aspect ratio randomly sampled in the range [0.1, 0.25]. The magnetic axis was parameterized as $R_{axis} = 1 + \xi \cos\phi$, $Z_{axis} = \xi \sin\phi$, where $\xi \in [0, 0.2]$ and $\phi$ is the toroidal angle in the cylindrical coordinates. The toroidal magnetic flux of last closed flux surface was fixed as $\psi_{tm} = 0.04 \times 2\pi$ Wb. The $\iota$ profile was prescribed by $\iota(\rho) = \iota_0 + 0.1\rho^2$, where $\iota_0$ is randomly selected from [0.35, 1.85], and $\rho$ is the square root of normalized toroidal flux, $\rho = \sqrt{\psi_t/\psi_{tm}}$. The pressure profile was prescribed by $p(\rho) = \left(\left(-0.5\tanh\left((\rho - p_1)/p_2\right) - 0.5\right) \times p_3 + 1\right) \times p_0$, with $p_0$ selected in the range $[10^3, 3\times10^4]$ Pa, $p_1$ selected in the range [0.2, 0.8], $p_2$ selected from [0.2, 0.5], and $p_3$ selected in the range [0.2, 0.95]. Then each randomly generated configuration was optimized by minimizing the 'two-term' quasi-symmetry error[43]

$$\hat{f}_C = \langle |(M\iota - N)(\mathbf{B} \times \nabla\psi) \cdot \nabla B - (MG + NI)\mathbf{B} \cdot \nabla B|\rangle / \langle B \rangle^3,$$

where $M = 1$, $N = 4$. $\psi$ is the toroidal flux function. $I$ and $G$ are the poloidal and toroidal components of the covariant form of $\mathbf{B}$ using Boozer coordinates. $I = \mathbf{B} \cdot \partial\mathbf{r}/\partial\theta_B$, $G = \mathbf{B} \cdot \partial\mathbf{r}/\partial\zeta_B$, where $\theta_B$ and $\zeta_B$ are, respectively, the poloidal and toroidal Boozer angle. The bracket $\langle \quad \rangle$ means volume averaging. During the optimization, the inverse aspect ratio was constrained to [0.1, 1/3], and the maximum elongation of all toroidal angles was limited to [1, 3]. The toroidal flux at boundary, the pressure profile, and the averaged radial position of boundary were fixed. In addition, the force balance of these equilibria was required by minimizing the relative force balance error,

$$|F| = |J \times B - \nabla p| / \langle |\nabla B|^2/(2\mu_0) \rangle$$

In order to highlight the geometrical effect on zonal flows and turbulence, two groups of datasets with fixed $\iota_0 = 0.55$ and $\iota_0 = 1.35$ were also generated, with all other settings the same as the random-$\iota_0$ dataset. Since our purpose is to find the hidden pattern of QS geometries, the obtained configurations with $\hat{f}_C > 0.1$ or $\langle |F| \rangle > 0.1$ were discarded, while the ones with good symmetry were kept in the dataset. The QS error distribution is shown in Fig.1. We have generated 2995 equilibria with random $\iota_0$, 5197 equilibria with $\iota_0 = 0.55$, and 5757 equilibria with $\iota_0 = 1.35$. Each equilibrium is represented by the coefficients $(R_{lmn}, Z_{lmn})$, which describe the flux surface shape,

$$R(\rho,\theta,\phi) = \sum_{lmn} R_{lmn} \mathcal{Z}_l^m(\rho,\theta)\mathcal{F}^n(\phi)$$

$$Z(\rho,\theta,\phi) = \sum_{lmn} Z_{lmn} \mathcal{Z}_l^m(\rho,\theta)\mathcal{F}^n(\phi),$$

where $\mathcal{Z}_l^m$ are the Fourier-Zernike basis functions and $\mathcal{F}^n$ are the Fourier basis functions. $(l, m, n)$ stand for the order of basis function in radial, poloidal and toroidal directions. In the current dataset, the basis functions are truncated at maximum order 8 in radial, poloidal, and toroidal directions. The 'ANSI' indexing for Zernike polynomials was chosen, and considering the stellarator symmetry, there are 385 $R$ coefficients and 380 $Z$ coefficients. In other words, the design space of these geometries has a dimension of 765.

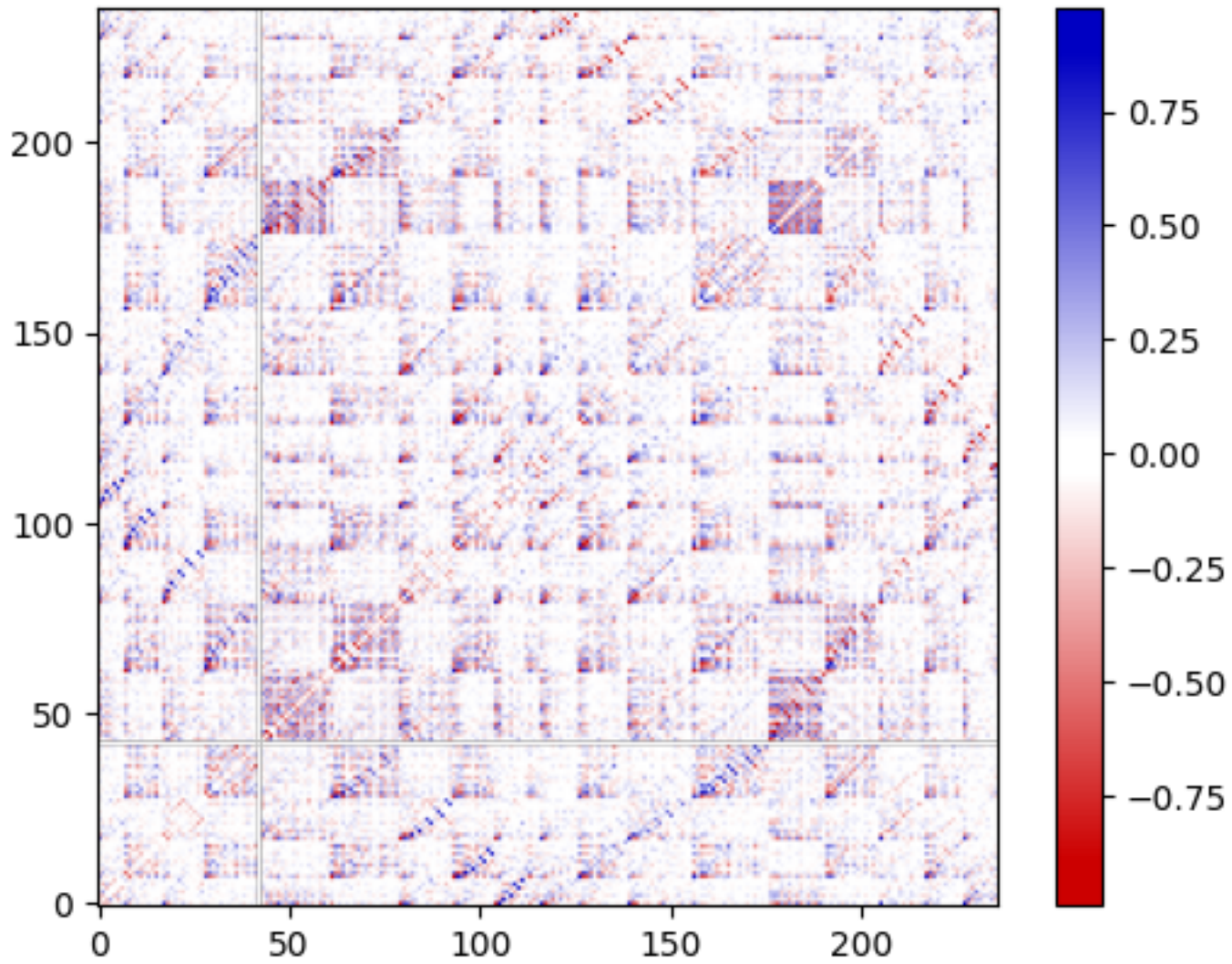

*Figure 2 The Pearson coefficient of the 250 (R,Z) coefficients with largest averaged absolute value.*

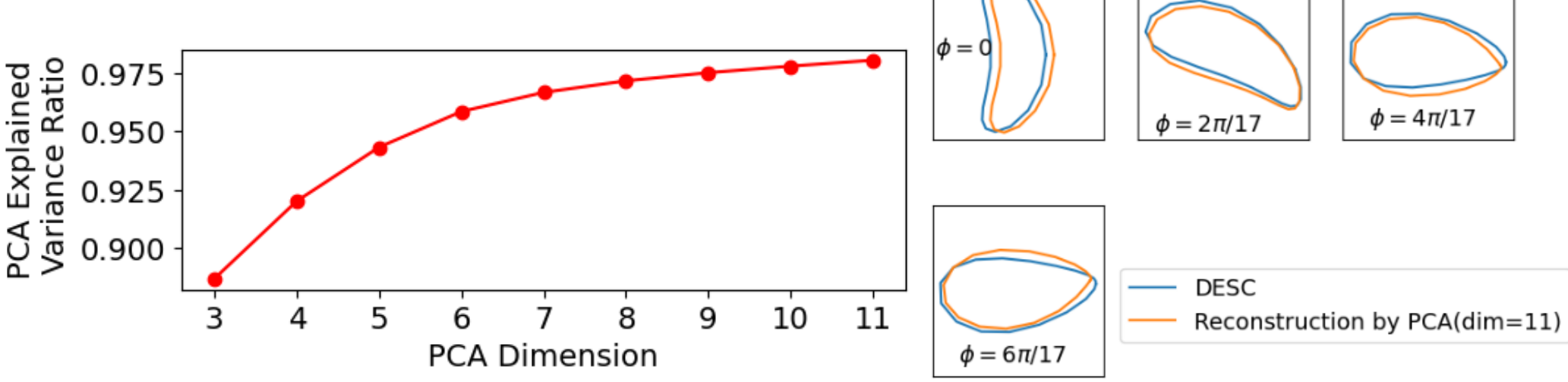

*Figure 3 Reconstruction by PCA. (Left) Dependence of EVR of all (R, Z) coefficients on the PCA dimensions. (Right) A example of the reconstructed last closed flux surface with 11-dimensional PCA compared with the ground truth data generated by DESC.*

We first evaluate the linear properties of the data distribution. The Pearson correlation coefficients[44] were calculated for the top 250 coefficients with the largest mean absolute values in Scipy[45] to show their linear relations. It is shown that largest coefficients have a mean correlation around $\pm 0.5$, with the largest magnitudes up to $\pm 0.8$, meaning a medium to strong linear relations. Note that the Pearson correlation only shows the linear relation among the parameters, which suggests the coefficients can be subject to even stronger nonlinear constraints, and a significant dimensionality reduction is feasible. Principal Component Analysis (PCA)[46], [47] was applied to evaluate the linear dimensionality reduction. The Explained Variance Ratio (EVR) is the ratio of total data variance in the reduced low dimensional space to the original data variance. When EVR approaches 1, it means most of the original data distribution feature can be preserved after the reduction. Fig.3 shows that EVR almost reaches unity in an 11-dimensional subspace. Using this 11-d space, the (R, Z) coefficients and the associated flux surface shapes can be reconstructed. Although the reconstructed flux surface shapes resemble the original optimized geometries from DESC, the detailed calculations show substantial deviation in |B| and the associated QS error. Furthermore, the 11-d space remains too large for efficient generation of simulation data and the training of surrogate model.

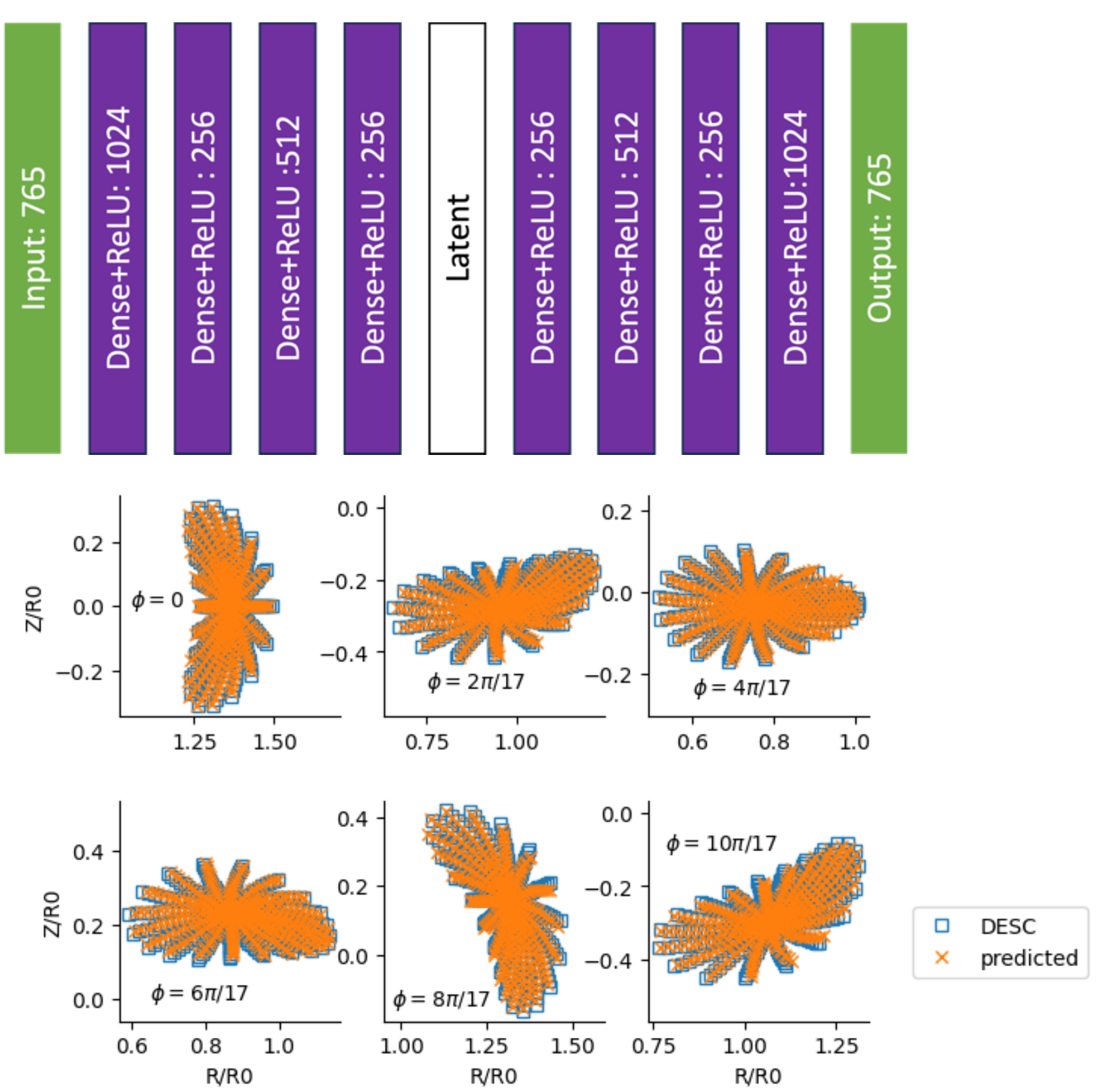

*Figure 4 (a) Autoencoder architecture. (b) Comparison of collocation point positions calculated from DESC coefficients and reconstructed coefficients at different toroidal angles.*

To achieve further dimensionality reduction by taking into account nonlinear relations, we constructed a fully connected autoencoder neural network. The input and the output consist of the 765-dimensional (R, Z) coefficient vector (noted by $\vec{C}$). The encoder and decoder each contain four hidden layers with the ReLU activation function. A low-dimensional latent layer is located at the bottleneck. The autoencoder structure is shown as Fig. 4(a). The primary loss term is the mean-squared reconstruction error of the coefficients, $L_{MSE} = \sqrt{\sum_{i=1,765}\left(C_{i,output} - C_{i,input}\right)^2}/765$. That is, the autoencoder is trained to ensure the coefficients to be effectively compressed from the original 765-d space to the low dimensional latent space. To improve the geometric reconstruction accuracy rather than just the coefficients, we incorporated an additional loss term based on collocation points. Specifically, 20x16x17 uniformly distributed grid points in $(\rho, \theta, \phi)$ space were used to calculate the value of $(R, Z, \phi)$ in the cylindrical coordinates by substituting the ground truth (R, Z) coefficients and reconstructed coefficients in Eq (1). The average distance, which was defined as collocation point loss, was included in the loss function, $L_{points} = \sum_{k=1,5440}\sqrt{\left(R_{k,DESC} - R_{k,pred}\right)^2 + \left(Z_{k,DESC} - Z_{k,pred}\right)^2}/5440$. The collocation points calculated from DESC coefficients and from the reconstructed coefficients can be seen from Fig. 4(b). The final loss function is $L = L_{MSE} + wL_{points}$. In the first 20 epochs during the training, only the coefficient loss was used (w=0), which makes the training process quickly converges. Subsequently, $w$ was set to be 0.25, and the collocation point loss was introduced to improve the accuracy of the geometric reconstruction.

## 3. Results

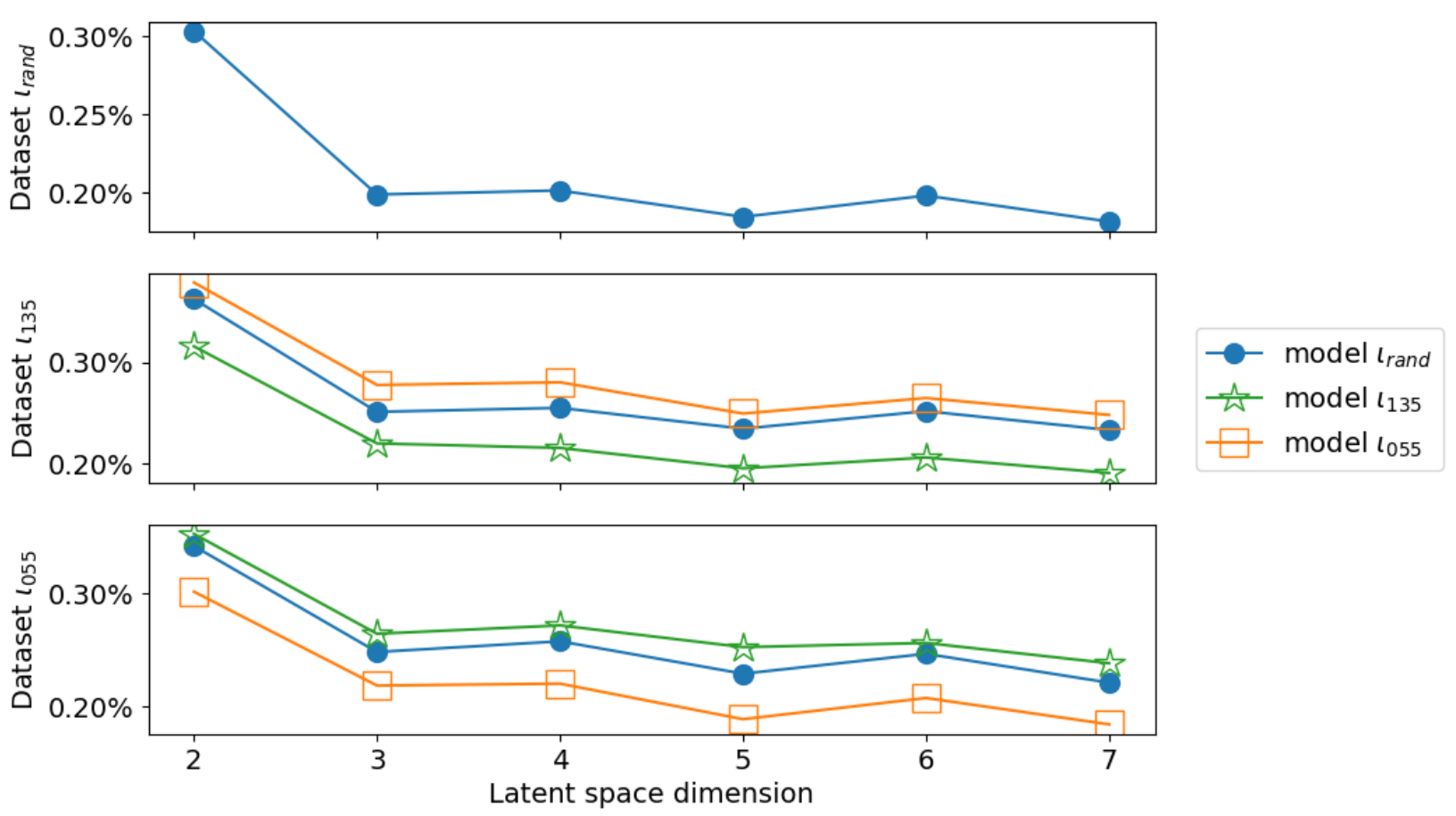

*Figure 5 The reconstruction error dependence on latent space dimensions. In the lower two panels, the models trained on different groups of data are tested.*

The latent space dimension was varied, and the reconstruction error $L_{MSE}$ on the test dataset was calculated to determine the minimum intrinsic dimensionality of the dataset. As shown in Fig. 5, the reconstruction error increases sharply when the latent dimension is smaller than 3, indicating that the dataset lies approximately on a three-dimensional nonlinear manifold. Note that the models trained on fixed or random $\iota_0$ datasets generalize well across different $\iota_0$ values, suggesting that the QH geometric distribution is largely independent of $\iota_0$. An example of the reconstructed equilibrium can be seen in Fig. 6. The left panel shows the reconstruction of $(R, Z)$ coefficients, and the right panel shows the reconstruction of the last closed flux surface. Compared with PCA, the autoencoder achieves significantly higher geometric reconstruction accuracy. Another test shows that removing the collocation points loss degrades flux surface reconstruction, although the coefficients reconstruction stays relatively accurate. On the other hand, using only the collocation loss prevents the convergence of the training process of the autoencoder.

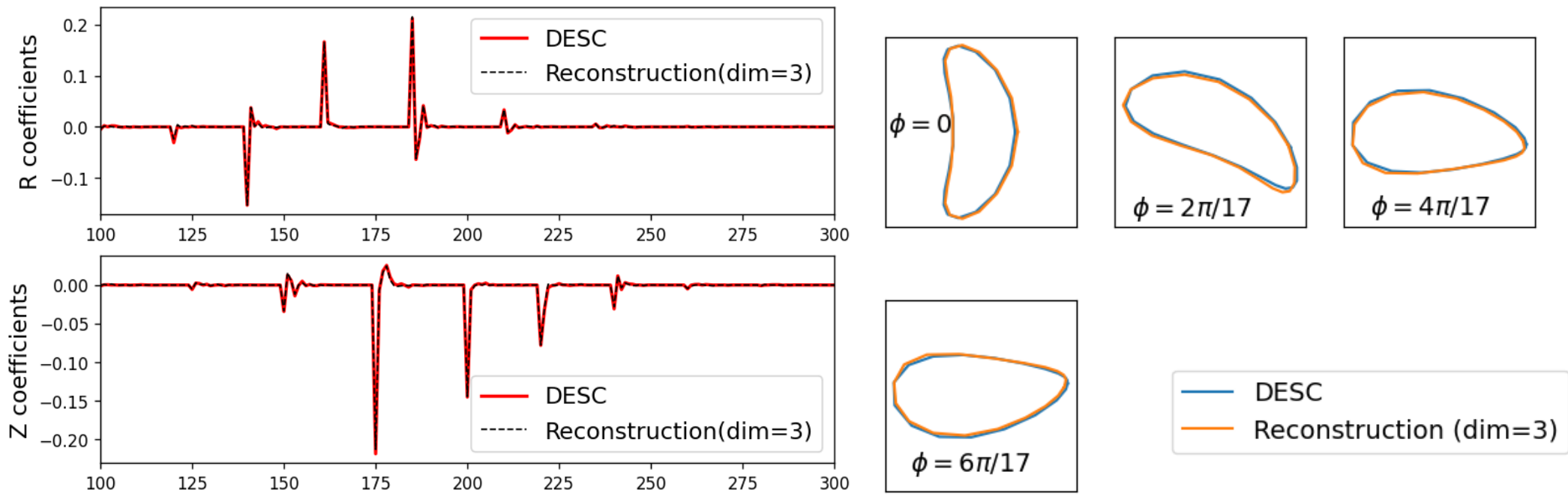

*Figure 6 An example of autoencoder reconstruction (latent dim=3) compared to the optimized RH geometry. (left panel) (R, Z) coefficients. the x-axes are the index of the modes, The coefficients too close to 0 are truncated for clear view. (right panel) Reconstructed last closed flux surface shape.*

The results above mean that the flux surface information can be compressed to very few parameters, which are the coordinates of the latent space. Together with the $\iota$ profile, the whole equilibrium and the physical quantities determined by the geometric parameters can be inferred from the latent space coordinates. The linear zonal flow (ZF) residual level is one important example of these physical quantities. The study of turbulent transport on different stellarator configurations showed that the ZF residual level can play an important role in determining the nonlinear steady-state ZF level and hence the transport level. Generally, ZF residual level can be characterized by its long wavelength limit, i.e., the Rosenbluth-Hinton (RH) level. In this sense, finding the geometrical factors that affect the RH level and a way to quickly predict the RH level is very important to finding transport-optimized designs. For this purpose, we used the latent space representation to directly predict the volume averaged RH level. The pre-trained encoder obtained in the reconstruction task maps the coefficients into latent space, followed by 3 fully connected layers with the width (64, 64, 32) for the regression of the RH level. The inputs are still the 765-d coefficients vector. The ground-truth RH level for the training data is calculated using the formula in [41],

$$
\begin{aligned}
RH(\psi) &= \frac{1}{1+\Lambda_1/\Lambda_0}, \\
\Lambda_1 &= \frac{m_i n_i (G+NI)^2}{B_0^2(\iota - N)^2}\left[1.6\epsilon^{3/2} + \mathcal{O}(\epsilon^2)\right], \\
\Lambda_0 &= m_i n_i \langle |\nabla\psi|^2/B^2 \rangle \, .
\end{aligned}
$$

Where $N = 4$ is the toroidal period in this dataset, and $\psi$ is the toroidal flux function. Here $\epsilon$ is a function of flux function which can be calculated from $\epsilon = (B_{max} - B_{min})/(B_{max} + B_{min})$ on the flux surface. Then the RH levels on different flux surfaces are averaged to get a single value as the ground truth value of the output. Since the $\iota$ profile will significantly affect the RH level, we only constructed the surrogate model for the two fixed $\iota_0$ datasets to highlight the geometrical effects. As a result, the predicted RH level agrees well with the theorical values, which can be seen from Fig.7. For a 3-d latent space, the average relative error is 15% for $\iota_0 = 1.35$ test dataset, and 2.5% for $\iota_0 = 0.55$ test dataset. If we increase the latent space to 7-d, the error reduces to 2.3% for $\iota_0 = 1.35$ test dataset, and 2.2% for $\iota_0 = 0.55$ test dataset.

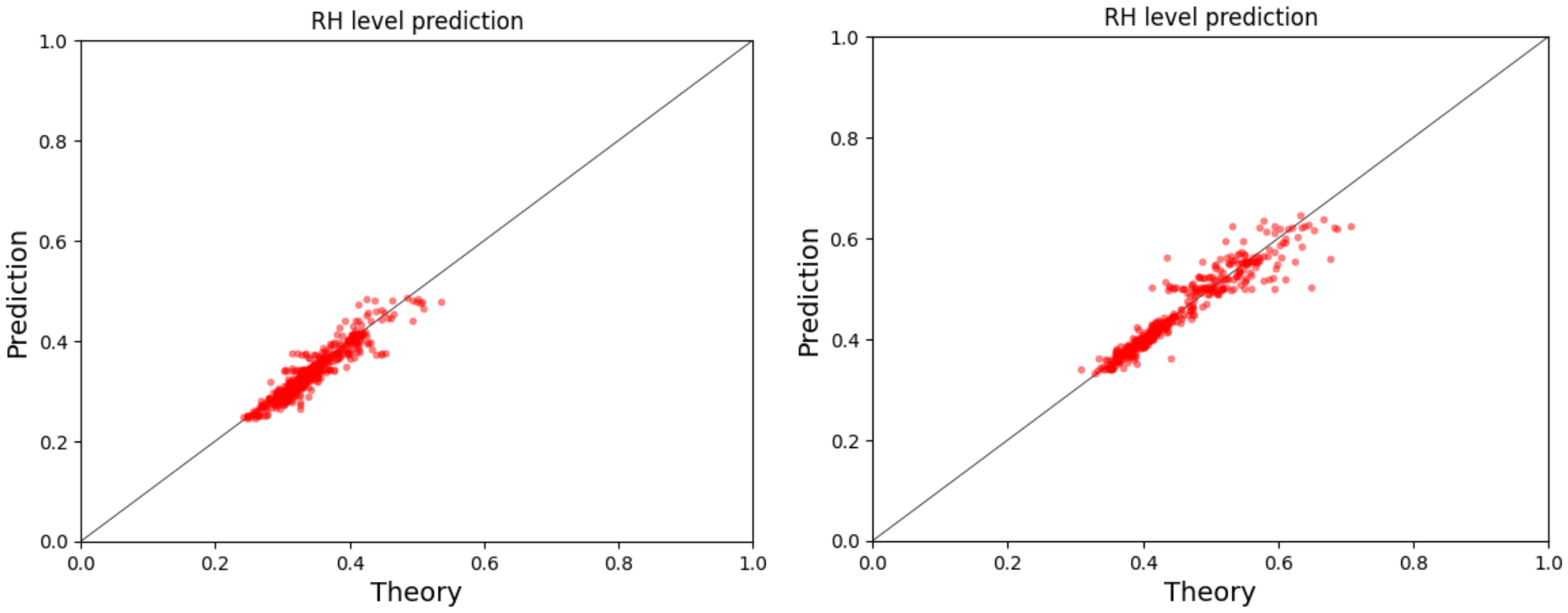

*Figure 7 Prediction of RH level compared to the theoretical values. The y=x line means perfect prediction. (Left)* $\iota_0 = 1.35$ *and (Right)* $\iota_0 = 0.55$.

The distribution of the RH levels of the training data with $\iota_0 = 1.35$ in 3-d latent space is shown in Fig.8. By analyzing the data in different regions, one important geometric factor is the aspect ratio, which directly influences both the $\nabla\psi/B$ term and the $\epsilon$ term in Eq. (1). However, even for equilibria with similar aspect ratio and elongation, significant variation in RH level is observed. This variation correlates with the magnetic axis excursion, defined as the amplitude of the magnetic axis displacement in the R-Z plane as $\phi$ changes. For configurations with similar aspect ratio and identical $\iota$ profile, equilibria with smaller axis excursion exhibit systematically higher RH levels. A comparison of such an equilibrium pair can be seen in Fig.8. This trend can be interpreted qualitatively using the first order near-axis expansion theory. The $\Lambda_1$ term in Eq. (1) is constant on the flux surface, so the RH level variation on the same flux-surface relies on the variation of $\Lambda_0$. It is found that, due to the large toroidal variation of the Jacobian, the flux-surface averaged $\Lambda_0$ is predominantly related to its value at $\zeta = 0$ (bean shape cross section). From [41], The RH level at $\zeta = 0$ can be written as

$$RH(\zeta = 0) = \frac{1}{1 + 1.6 / \left(\sqrt{\varepsilon} \mathcal{C} (\iota - N)^2\right)},$$
$$\mathcal{C} = \alpha / (\bar{\eta} R_a)^2, \alpha = \langle |\nabla \psi|^2 / B^2 \rangle,$$
$$\bar{\eta} = \sqrt{2\psi / B} / \varepsilon,$$

Where $R_a$ stands for the magnetic axis length over one toroidal period. For equilibria with similar aspect ratio and identical $\iota$ profile, the factors $\alpha$ and $\bar{\eta}$ factor vary weakly. Consequently, the magnetic axis length $R_a$ becomes the dominant geometric factor controlling $\mathcal{C}$, and hence the RH level. Smaller axis excursion corresponds to shorter magnetic axis length, which increases $\mathcal{C}$ and therefore increases the RH level. This interpretation provides a clear physical explanation for the latent space structure observed in the data. The autoencoder has effectively identified a low-dimensional representation in which the dominant geometric control parameters for zonal flow residuals are embedded.

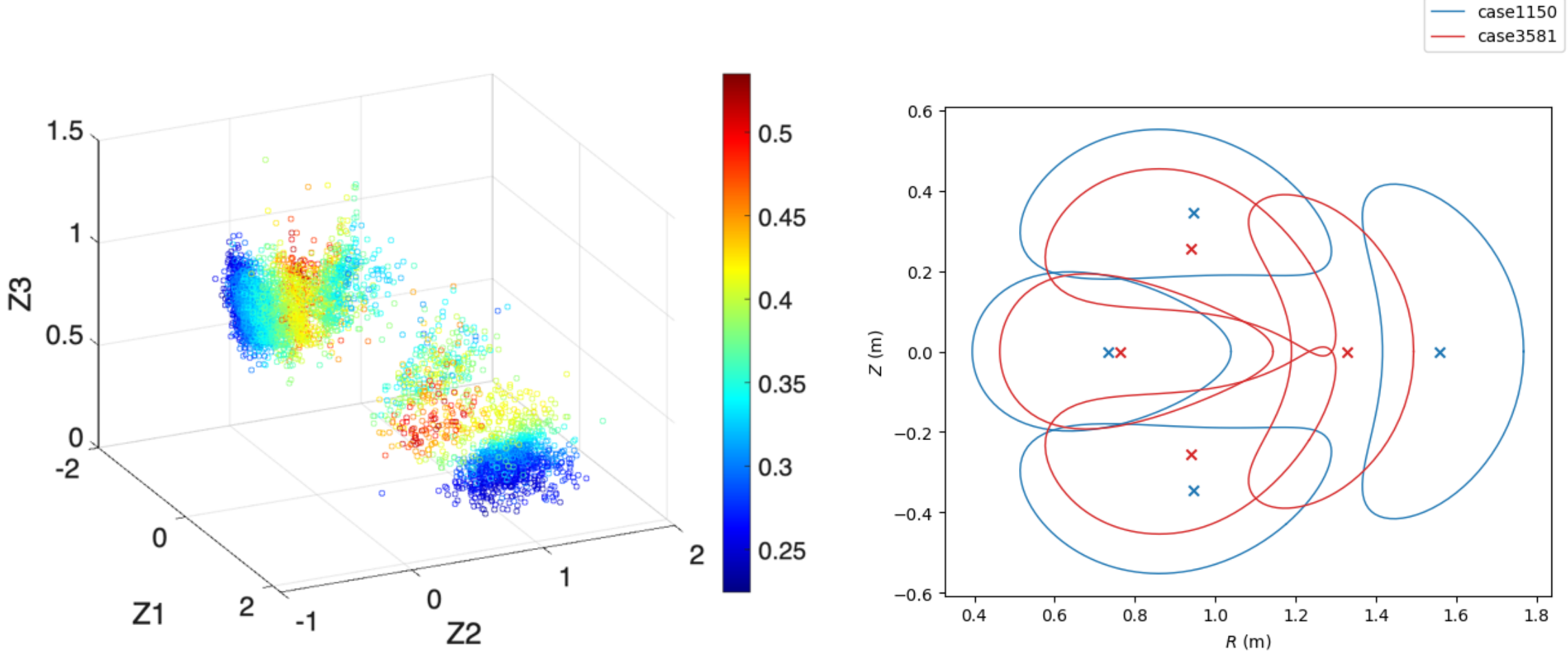

*Figure 8 (left) The theoretical RH levels of* $\iota_0 = 1.35$ *training data in the 3-d latent space. (right) A comparison of two equilibria with similar aspect ratio and elongations, but different axis excursions. The mean RH level over* $\rho$ *for case3581 is 0.42, while the mean RH for case1150 is 0.23.*

To systematically evaluate the turbulent transport properties across the generated QH stellarator database, more than 15,000 global GTC gyrokinetic simulations were performed. All equilibria were resized to isolate the geometric effects. The minor radius was set to $a = 0.8$ m, and the averaged on-axis B strength was fixed at $B_a = 4.9$ T. The ion and electron density profile are set to be uniform with $n_0 = 8.3 \times 10^{13}$ cm$^{-3}$. The ion and electron temperature profiles were set to have the largest gradient near $\rho = 0.5$ with the minimum scale length being $a/L_{Ti} = 4.7$. The on-axis temperatures were set to $T_{i0} = T_{e0} = 40$ keV, corresponding to $a/\rho_i \approx 194$ on axis and 220 at the lowest $L_{Ti}$ location. This set of parameters can destabilize the ITG mode in most cases. The resulting steady-state turbulent transport varied significantly across configurations. Excluding the cases in which ITG mode was linearly stable, the steady-state ion heat conductivity ranged from approximately $10^{-4}\chi_{gB}$ to $10^{-1}\chi_{gB}$, where $\chi_{gB} = T_{e0}/eB_a * \rho^*$ is the gyro-Bohm unit, $\rho^* = \rho_i/a$ at the mode location (lowest $L_{Ti}$). This broad distribution suggests the strong sensitivity of turbulent transport to magnetic geometry within the QH parameter space. A systematic trend is observed that the configurations with larger $|\iota - N|$ generally show lower ion heat conductivity, consistent with the previous conclusion in [29], [41]. When the steady-state transport levels are projected onto the three-dimensional latent space identified by the autoencoder, recognizable structure emerges. Although the clustering pattern is less clear than in the RH level analysis, regions associated with reduced transport can still be identified. Taking advantage of this property, we constructed a simple generative model in latent space to explore the transport-optimized designs. A Gaussian mixture model was fitted to equilibria satisfying $\chi_i < 5 \times 10^{-3} \chi_{gB}$. New latent coordinates were sampled from the fitted distribution in the 3-d latent space. Then the data were mapped back to the full geometric coefficients using the trained decoder. It is confirmed by the GTC simulations that the new samples do have a distribution with a lower average transport level compared to the randomly generated original dataset, which is particularly pronounced for the $\iota_0 = 0.55$ dataset.

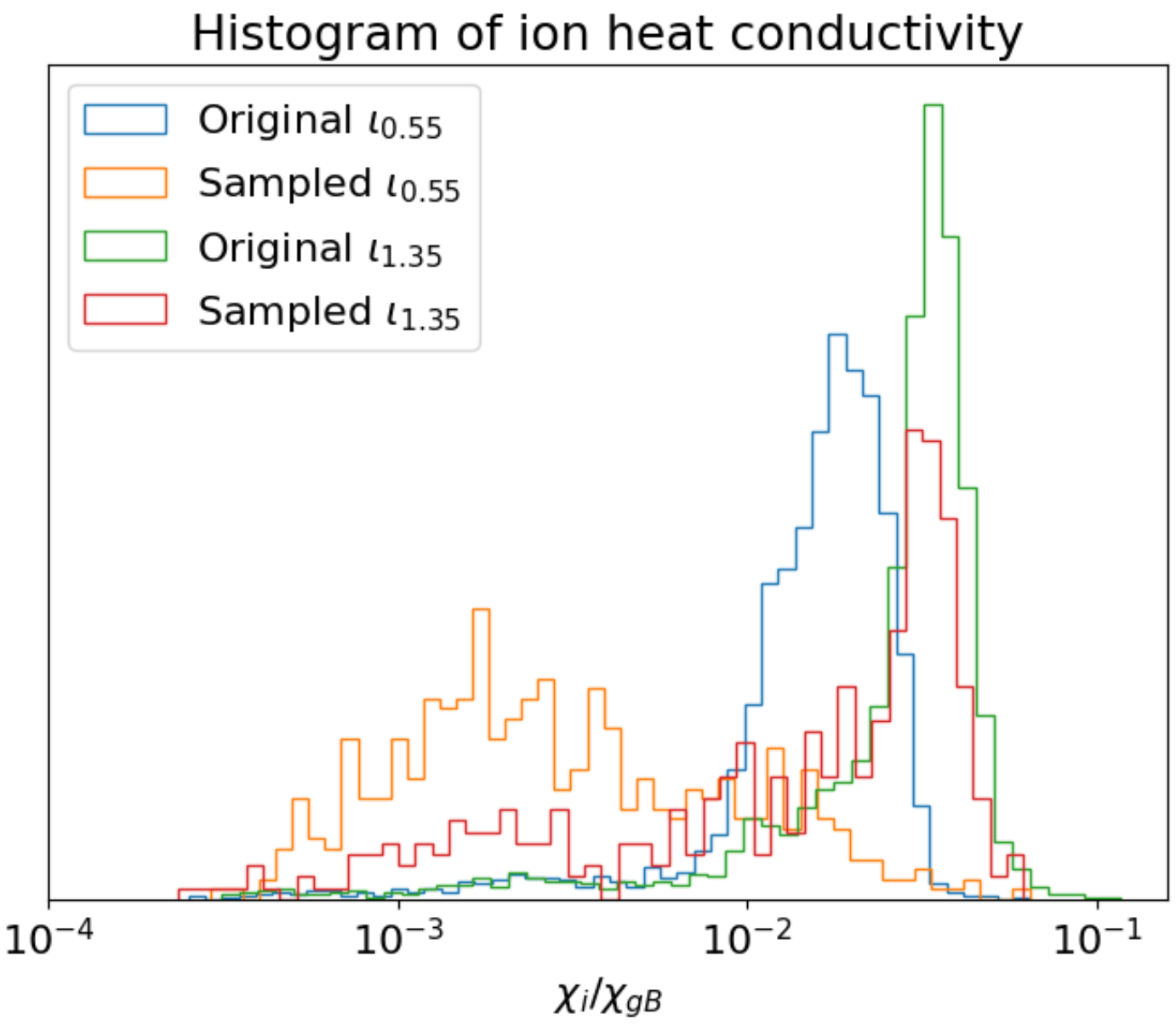

*Figure 9 Distribution of the ion heat conductivity of different datasets.*

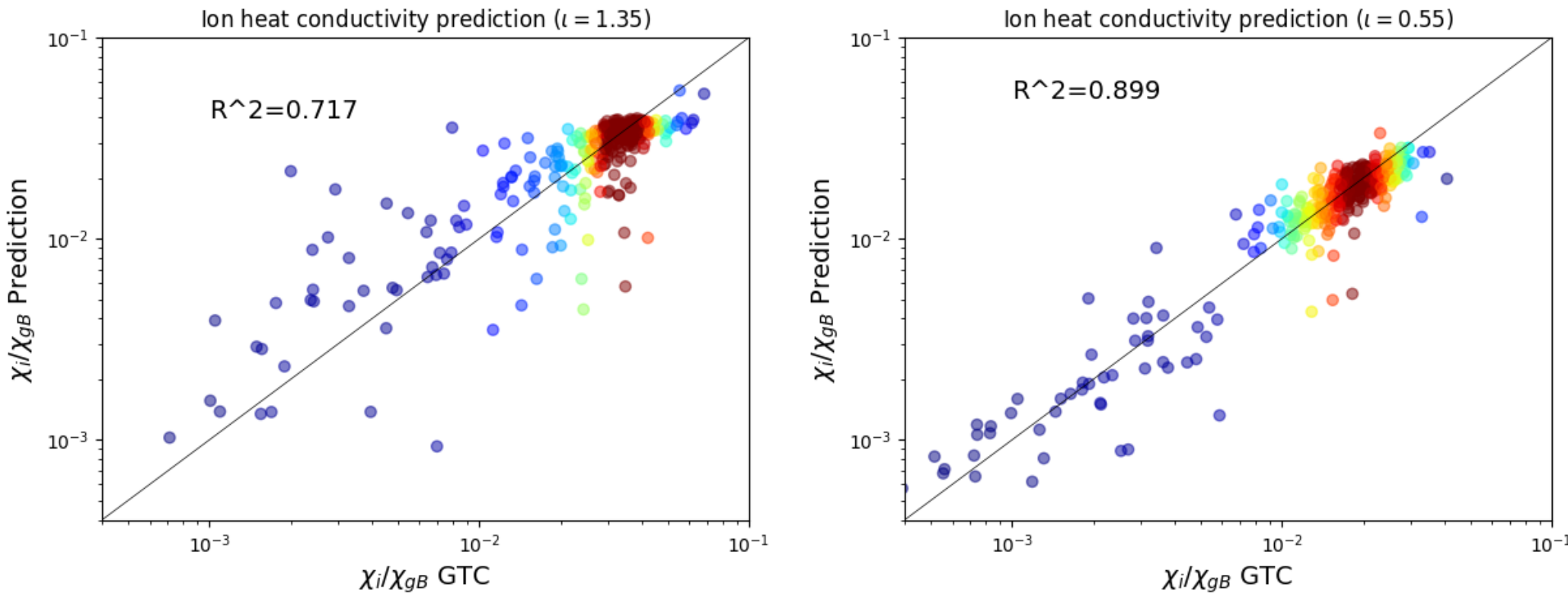

*Figure 10 Prediction of volume averaged ion heat flux driven by ITG compared with GTC simulations.*

Beyond qualitative identification of low-transport regions, a quantitative surrogate model was developed to predict the steady-state transport level for QH geometries. The neural network architecture is the same as that used for the RH level prediction, with the output replaced by the steady state $\chi_i/\chi_{gB}$ value for each case. If the encoder weights are frozen and only the regression layers are trained, like the RH case, the prediction accuracy of $\chi_i$ remains poor. This indicates that the latent space optimized for geometric reconstruction does not necessarily preserve the geometric features most relevant for nonlinear transport. The best performance was obtained through a two-stage training strategy. First, the regression layers are trained for 30 epochs while keeping the encoder fixed. Then the encoder is fine-tuned jointly with the regression layers using a reduced learning rate. This fine-tuning allows the latent representation to adjust slightly so that the transport-relevant geometric features are better resolved. After fine-tuning, the predicted transport levels agree reasonably with the GTC gyrokinetic results, particularly for the $\iota_0 = 0.55$ dataset. The comparison between the predicted and simulated ion heat conductivity is shown in Fig. 10. For $\iota_0 = 0.55$, the coefficient of determination $R^2$ approaches 0.9, indicating strong predictive capability. In contrast, the $\iota_0 = 1.35$ dataset exhibits lower prediction accuracy and larger scatter. The difference between the two datasets is also reflected in the $\chi_i$ of the sampled data in Fig.9, where it is easier to sample out the low transport cases for $\iota_0 = 0.55$ dataset. These results highlight two properties of the distribution of transport levels. First, while the geometric manifold is intrinsically low-dimensional, turbulent transport depends on more subtle geometric features than those required for accurate flux surface reconstruction. Second, moderate fine-tuning of latent representation significantly improves transport prediction without sacrificing geometric interpretability.

## 4. Conclusion and future work

This study demonstrates that QH stellarator geometries exhibit intrinsic dimensionality as low as 3, and hence it is possible to construct a surrogate model to predict the global features of the stellarator with ~10k data points due to this feature. The RH zonal flow residual can be predicted accurately from the coordinates of the latent space, which may have an impact on future stellarator designs. A direct relationship between the RH level and the magnetic axis excursion is identified and supported by the near-axis expansion theory. Furthermore, the large-scale GTC global gyrokinetic turbulence simulation confirms significant variation in transport properties across different QH geometries. This dataset can be very helpful for future optimized stellarator designs. For example, the latent space together with the simulation data can be used to directly sample new geometries with low transport level while keeping the QH symmetry. Although the original latent space trained from reconstruction is not appropriate for the quantitative prediction of transport level, it can be fine-tuned to improve the prediction accuracy. Although the prediction is not perfect, this work can be a proof of concept using dimensionality reduction and surrogate models to optimize the global features of stellarators.

Future work will focus on improving the prediction accuracy of the transport level through simulation data cleaning, hyperparameter search, and ensemble learning with diverse network architectures. Integration of the transport surrogate model into the optimization framework will allow simultaneous optimization of quasi-symmetry, turbulence, and other properties. Extension of the methodology to QA and QI geometries will also be investigated. Finally, we will broaden the usage of the autoencoder in stellarator optimization, such as geometry generation and PINN from the latent space.

## Acknowledgement

This work is supported by the US Department of Energy (DOE) SciDAC project HifiStell. An award of computer time was provided by the U.S. Department of Energy's (DOE) Innovative and Novel Computational Impact on Theory and Experiment (INCITE) Program. This research used supporting resources at the Argonne and the Oak Ridge Leadership Computing Facilities. The Argonne Leadership Computing Facility at Argonne National Laboratory is supported by the Office of Science of the U.S. DOE under Contract No. DE-AC02-06CH11357. The Oak Ridge Leadership Computing Facility at the Oak Ridge National Laboratory is supported by the Office of Science of the U.S. DOE under Contract No. DE-AC05-00OR22725. This research used resources of the National Energy Research Scientific Computing Center (NERSC), a Department of Energy User Facility using NERSC award FES-ERCAP0037742 and DDR-ERCAP0035957. This report was prepared as an account of work sponsored by an agency of the United States Government. Neither the United States Government nor any agency thereof, nor any of their employees, makes any warranty, express or implied, or assumes any legal liability or responsibility for the accuracy, completeness, or usefulness of any information, apparatus, product, or process disclosed, or represents that its use would not infringe privately owned rights. Reference herein to any specific commercial product, process, or service by trade name, trademark, manufacturer, or otherwise does not necessarily constitute or imply its endorsement, recommendation, or favoring by the United States Government or any agency thereof. The views and opinions of authors expressed herein do not necessarily state or reflect those of the United States Government or any agency thereof.